\documentstyle[aps,prl,epsf]{revtex} 
\draft
\begin{document}
% Macros for the various macro package names, etc.
\def\SNG{{\em Physical Review Style and Notation Guide}}
\def\LUG {{\em \LaTeX{} User's Guide \& Reference Manual}}
\def\btt#1{{\tt$\backslash$\string#1}}%
\def\REVTeX{REV\TeX}
\def\AmS{{\protect\the\textfont2
        A\kern-.1667em\lower.5ex\hbox{M}\kern-.125emS}}
\def\AmSLaTeX{\AmS-\LaTeX}
\def\BibTeX{\rm B{\sc ib}\TeX}
%\makeatletter
%\tighten
%\twocolumn[\hsize\textwidth\columnwidth\hsize\csname@twocolumnfalse%
%\endcsname
\title{Quantum ferromagnetic transition in disordered itinerant electron systems} 
\author{D.Belitz}
\address{Department of Physics and Materials Science Institute,%\\ 
University of Oregon,%\\
Eugene, OR 97403}
\author{T.R.Kirkpatrick}
\address{Institute for Physical Science and Technology, and Department of Physics\\
University of Maryland, College Park, MD 20742}
\date{\today}
\maketitle
\begin{abstract}
An effective field theory is derived for the ferromagnetic transition
of diffusive electrons at $T=0$. The static disorder which leads to 
diffusive electron
dynamics induces an effective long-range
interaction between the spins of the form $1/r^{2d-2}$. This leads to
unusual scaling behavior at the quantum critical point, which is
determined exactly. The crossover from this quantum fixed point to
the classical Heisenberg fixed point should be observable in 
ferromagnetic materials with low Curie temperatures.
\end{abstract}
%]
%\narrowtext
     
Quantum phase transitions have attracted substantial interest in recent
years. They occur at zero temperature ($T=0$) as a
function of some non-thermal control parameter, and the relevant
fluctuations are of quantum rather than of thermal nature.
If the quantum phase transition has a classical analogue at finite
temperature, then the former tends to be simpler than the latter
in the physical dimension $d=3$. For instance,
the ferromagnetic-to-paramagnetic transition
of itinerant electrons at $T=0$ as a function
of the exchange interaction strength $\Gamma_t$, has a
mean-field like critical behavior in $d=3$ \cite{Hertz}, 
since the coupling of statics and dynamics
that is inherent to quantum statistics problems effectively increases
the dimensionality of the system from $d$ to $d+z$ with $z$ the
dynamical critical exponent. In the case of clean quantum ferromagnets,
$z=3$, which reduces the upper critical dimension
from $d_c^+ = 4$ in the classical case to $d_c^+=1$\cite{Hertz}.

If one adds quenched disorder in the form of static, nonmagnetic scatterers
to the problem, then the resulting diffusive
dynamics of the electrons changes the dynamical exponent $z$.
Hertz\cite{Hertz} has proposed that
$z=4$ in the disordered case, which would imply that $d_c^+=0$,
that the critical behavior is otherwise unchanged, and that in particular
the correlation length exponent $\nu=1/2$ for all $d>0$. It is now known
that this cannot be true for $d<4$, since $\nu=1/2$ would violate the
Harris criterion, which states that any fixed point (FP) with
$\nu < 2/d$ is unstable with respect to quenched disorder\cite{Harris,Millis}. 
In this Letter we show that Hertz's proposition is true
only for $d>6$, while for $2<d<6$ the disordered itinerant 
quantum ferromagnet shows nontrivial critical behavior which we 
determine exactly. The critical behavior is dominated by the strong 
static correlations that are induced by the disorder, not by the 
quantum fluctuations.
Indeed, we find $z=d$ for $2<d<4$, which leads to $d_c^+=2$, 
so that we work above the upper critical dimension. However, the disorder
leads to additional soft modes that lead to an
effective long-range interaction between the spins, and that are
responsible for the nontrivial critical behavior.

Our results can be summarized as follows. For the
correlation length exponent
$\nu$, the order parameter susceptibility exponent $\eta$, and the
dynamical exponent $z$, we find for $2<d<4$,
\begin{equation}
\nu = 1/(d-2)\quad,\quad \eta = 4-d\quad,\quad z=d\quad,
\label{eq:1}
\end{equation}
and $\nu=1/2$, $\eta=0$, $z=4$ for $d>4$. These exponents
`lock into' mean-field values at $d=4$, but have nontrivial values for
$d<4$. The magnetization, $m$, at
$T=0$ in a magnetic field $H$ is given by the equation of state
\begin{mathletters}
\label{eqs:2}
\begin{equation}
t\,m + v\,m^{d/2} + u\,m^3 = H\quad,
\label{eq:2a}
\end{equation}
where $t$ is the dimensionless distance from the critical point, and
$u$ and $v$ are finite numbers. From Eq.\ (\ref{eq:2a})
one obtains the critical exponents $\beta$ and $\delta$, defined by
$m \sim t^{\beta}$ and $m \sim H^{1/\delta}$, respectively, at $T=0$, as
\begin{equation}
\beta = 2/(d-2)\quad,\quad\delta = d/2\quad,
\label{eq:2b}
\end{equation}
\end{mathletters}%
for $2<d<6$, and $\beta=1/2$, $\delta=3$ for $d>6$. These exponents
`lock into' their mean-field values only at $d=6$.
Equation (\ref{eq:2a}) applies strictly to
$T=0$, so $t$ can be changed by changing the disorder or $\Gamma_t$, 
but {\it not} the temperature. At finite temperature,
we find that $m$ obeys a homogeneity law,
\begin{equation}
m(t,T,H) = b^{-\beta/\nu} m(tb^{1/\nu}, Tb^{\phi/\nu}, Hb^{\delta\beta/\nu})
                                                                    \quad,
\label{eq:3}
\end{equation}
where $b$ is an arbitrary scale factor.
$\phi = 2\nu$ for all $d>2$ is a crossover exponent that describes the 
crossover from the quantum FP to the classical Heisenberg FP. The behavior
of the magnetic susceptibility, $\chi_m$, follows from Eq.\ (\ref{eq:3})
by differentiation with respect to $H$. In particular, the exponent $\gamma$,
defined by $\chi_m \sim t^{-\gamma}$ at $T=H=0$, is 
$\gamma=\beta (\delta -1) =1$ for all $d>2$.
Notice that the temperature dependence of the magnetization is {\it not}
given by the dynamical exponent. $z$ controls the temperature dependence
of the specific heat coefficient, $\gamma_V = c_V/T$, however, which
has a scale dimension of zero for all $d$, and logarithmic corrections 
to scaling for all $d<4$,
\begin{equation}
\gamma_V(t,T,H) = \Theta(4-d)\,\ln b +
           \gamma_V(tb^{1/\nu}, Tb^z, Hb^{\delta\beta/\nu})\quad .
\label{eq:4}
\end{equation}

Equations (\ref{eq:1}) - (\ref{eq:4}) represent the exact critical
behavior of disordered itinerant quantum Heisenberg ferromagnets in all
dimensions $d>2$ with the exception of $d=4$. 
There are several remarkable aspects of these results. First,
we are able to obtain the critical behavior in $d=3$ exactly, yet it is not
mean-field like. The exactness is due to the fact that we work above
the upper critical dimension $d_c^+ = 2$. The nontrivial exponents are
due to the presence of noncritical soft modes, or spectator modes, which 
lead, e.g., to the
unusual term $\sim v$ in Eq.\ (\ref{eq:2a}). These points will become
clearer in connection with Eqs.\ (\ref{eq:9}), and
(\ref{eq:10}) below.
Another point of interest is the appearance of a logarithmic correction
to scaling in Eq.\ (\ref{eq:4}).
In a renormalization group context, logarithmic anomalies can arise
either due to marginal operators, or if a set of scale dimensions
fulfills some resonance condition \cite{Wegner}. In the present case the
logarithm can be traced to the condition $z = d$, which is fulfilled
{\it for all} $2<d<4$. The behavior in $d=4$ is more complicated.\cite{ustbp}.

The above values for the critical exponents $\beta$ and $\delta$ are
very different from those for classical Heisenberg magnets \cite{Ma}.
Consequently, we expect a pronounced crossover from quantum to classical
critical behavior as the ferromagnetic transition is approached at low
temperatures. Our results can be experimentally checked
in alloys such as Ni-Al or Ni-Ga \cite{Mott}. Especially the small value
of $\delta = 3/2$ in $d=3$ should be a prominent feature that could be
observed for a fixed alloy composition as a function of magnetic field. 
A more thorough discussion of experimental aspects of our theory, and a
complete account of technical details, will be given in Ref. \cite{ustbp}.

In the remainder of this paper we sketch the derivation of these results.
Let us consider a $d$-dimensional system ($d>2$) of interacting 
electrons in the presence of nonmagnetic quenched disorder. We
keep the disorder strength sufficiently small
for the system to be metallic\cite{locfootnote}. With increasing strength
of the exchange interaction
this system undergoes a phase transition from a paramagnetic metal to a
ferromagnetic metal. At $T=0$ this is the quantum phase transition we
are concerned with. At any $T>0$ the asymptotic critical behavior is
governed by a classical FP, but at low temperatures there
still exists a sizeable quantum critical region that is controlled by
the zero temperature FP. We proceed to derive an order parameter 
description of the quantum phase transition, along the lines of
Ref.\ \onlinecite{Hertz}. To this end, it is convenient to split off 
from the action the particle-hole spin-triplet channel, whose coupling constant
we denote by $\Gamma_t$. This four-fermion interaction
term we decouple by means of a Hubbard-Stratonovich (H-S) transformation with
a H-S field $M(q)$. All degrees of freedom
other than ${\bf M}$ are then integrated out. This procedure in
particular integrates out diffusive modes in the spin-triplet channel.
These are the soft `spectator' modes, which have important consequences
as we will see. We obtain the partition function $Z$ in the form
$Z = e^{-F_0/T} \int D[{\bf M}]\,\exp\bigl[-\Phi[{\bf M}]\bigr]$,
with $F_0$ the noncritical part of the free energy.
A formal expansion of the Landau-Ginzburg-Wilson (LGW) functional $\Phi$ in
powers of $M$ takes the form\cite{Hertz},
\begin{eqnarray}
\label{eq:7}
\Phi[{\bf M}] = \int_q{\bf M}(q) \bigl[1 - \Gamma_t\chi^{(2)} (q)
                \bigr]{\bf M}(-q)
              + {1\over V}\int_{q_1,q_2,q_3}
                \chi^{(4)}(q_1,q_2,q_3)\,
                  \bigl({\bf M}(q_1)\cdot{\bf M}(q_2)\bigr)
\nonumber\\
                  \times\bigl({\bf M}(q_3)\cdot{\bf M}(-q_1-q_2-q_3)\bigr)
              - \sqrt{V}\,H\,M_z(q=0)\:.
\end{eqnarray}
Here we are using a four-vector notation where
$q=({\bf q},\omega_n)$ comprises both a wavevector and a Matsubara
frequency, and $\int_q = \sum_{\bf q}\,T\sum_n$. $V$ is the system volume,
$H$ denotes a magnetic field in z-direction.
For notational simplicity we suppress the replica labels that are needed
to carry out the disorder average\cite{ustbp}.
$\chi^{(2)}$ and $\chi^{(4)}$ are disorder averaged correlation functions of a 
reference system of disordered electrons that has $\Gamma_t=0$\cite{Hertz}.
More generally, the coefficient of the term of
order ${\bf M}^n$ in the LGW functional is given in terms of an $n$-point
spin-density correlation function in the reference system, and in writing
Eq.\ (\ref{eq:7}) we have left out terms which can be shown to be
irrelevant for the critical behavior \cite{ustbp}. We will come back to
these omitted terms in the end.
The particle-hole spin-triplet interaction $\Gamma_t$ is missing in the bare
reference system, but it is generated by
perturbation theory already at the one-loop level.
The reference system then has
all of the characteristics of the full system,
except that it must not undergo a phase transition lest the separation of
modes that is implicit in our singling out the spin-triplet channel for the
H-S decoupling procedure break down.

$\chi^{(2)}(q)$ in Eq.\ (\ref{eq:7}) is the spin susceptibility of the 
reference system.  At small Matsubara frequency and 
wavenumber it has the characteristic diffusive form 
\cite{Forster},
\begin{equation}
\chi^{(2)}({\bf q},\omega_n) 
           = \chi_0({\bf q})[1 - \vert\omega_n\vert/D{\bf q}^2]\quad,
\label{eq:8}
\end{equation}
where $D$ is the spin diffusion constant of the reference system, and
$\chi_0({\bf q})$ is its static spin susceptibility. At $T=0$,
the latter can be shown to be a nonanalytic function of ${\bf q}$ of the form
$\chi_0({\bf q}\rightarrow 0) \sim {\rm const} - \vert{\bf q}\vert^{d-2}
- {\bf q}^2$, where we have omitted irrelevant prefactors \cite{chifootnote}. 
Choosing suitable units, the Gaussian part of $\Phi$ can then be written,
\begin{equation}
\Phi^{(2)}[{\bf M}] = \int_q {\bf M}(q)\bigl[t_0
                       + c_n\vert{\bf q}\vert^{d-2} + c_a{\bf q}^2 
                + c_d\vert\omega_n\vert/{\bf q}^2\bigr]\,{\bf M}(-q)\quad,
\label{eq:9}
\end{equation}
where $t_0 = 1 - \Gamma_t\chi^{(2)}({\bf q}\rightarrow 0,\omega_n = 0)$ 
is the bare distance
from the critical point, and $c_n$, $c_a$ and $c_d$ are constants. 

The coefficient $\chi^{(4)}$ in Eq.\ (\ref{eq:7}),
contrary to the case of a usual LGW functional, is in general not finite 
at zero frequencies and wavenumbers. 
$\chi^{(4)}$ is a nonlinear susceptibility in the reference ensemble.
Such a quantity one expects to be divergent, given the singularity in
the linear susceptibility $\chi_0({\bf q})$. Indeed,
standard perturbation
theory shows\cite{ustbp} that it is given schematically by
\begin{equation} 
\chi^{(4)} \sim {\rm const} + v\int_k \bigl[{\bf k}^2 
                            + \vert\omega_n\vert\bigr]^{-4} 
           \sim u +  v\vert{\bf p}\vert^{d-6}\ .  
\label{eq:10} 
\end{equation} 
Here $u$ and $v$ are finite numbers, and
we have cut off the singularity by means of a wave\-number
$\vert{\bf p}\vert$. The physical interpretation of this cutoff will
be dealt with below.
More generally, the coefficient of $\vert{\bf M}\vert^{2n}$ in $\Phi$
behaves like $\chi^{(2n)} \sim \vert{\bf p}\vert^{d+2-4n}$.
This implies that $\Phi$ contains a nonanalyticity
which in our expansion takes the form of a power series in 
$\vert{\bf M}\vert^2/\vert{\bf p}\vert^4$. 
The nonanalytic structure of the correlation functions $\chi^{(2n)}$
is a consequence of the diffusive spectator modes which we have integrated
out. Therefore the ${\bf M}$-field theory, rather than having a 
simple LGW form, is strongly nonlocal, which leads to unusual scaling behavior.

The functional $\Phi$ can be analyzed by using standard 
techniques \cite{Ma}. We look for a FP where $c_d$ and either $c_n$ 
(for $2<d<4$), or $c_a$ (for $d>4$) are not renormalized. This fixes the 
critical exponents $\eta$ and $z$. Power counting then reveals that both
$u$ and $v$ are irrelevant.
Several features of the critical
behavior follow immediately from this observation. The
critical exponents $\eta$ and $z$ are fixed by the choice of our FP, and
$\nu$ and $\gamma$ can be obtained by considering the
${\bf q}$-dependence of the Gaussian vertex, Eq.\ (\ref{eq:9}). We can thus
read Eq.\ (\ref{eq:1}) and the value of the exponent $\gamma$
directly off Eq.\ (\ref{eq:9}).

We determine the equation of state by taking
the term of order $\vert{\bf M}\vert^4$ in $\Phi$ into account.
To this end, we need to identify the scaling behavior of the cutoff
wavenumber ${\bf p}$. Since ${\bf p}$ cuts off a divergence stemming
from an integration over products of diffusion poles that is rendered
finite by spin-flip processes, in the presence
of an external magnetic field $B$ one would identify ${\bf p}^2$ with $B$.
Since a nonzero magnetization acts physically like a magnetic field, it
is therefore plausible that in the present context, and for scaling
purposes, ${\bf p}$ can be replaced by $m^{1/2}$. We have confirmed this
conclusion by means of a more technical analysis\cite{ustbp}.
The effective equation of state can therefore be written in the
form of Eq.\ (\ref{eq:2a}).
 
These results completely specify the critical behavior at $T=0$. Their
most interesting aspect is the nontrivial exponent values found for
$2<d<4$, which can nevertheless be determined exactly. The reason for this is
the $\vert{\bf q}\vert^{d-2}$-term in the Gaussian action, Eq.\ (\ref{eq:9}).
It reflects the fact that in an interacting electron system with quenched
disorder, static correlations between spins do not fall off exponentially with
distance, but only algebraically like $r^{-(2d-2)}$. This leads to a
long-range interaction that
falls off like $1/r^{2d-2}$, see Eq.\ (\ref{eq:9}). The critical
behavior of {\it classical} Heisenberg magnets with such a long-range
interaction has been studied before \cite{FisherMaNickel}.

We now turn to the
$T$-dependences of the specific heat, $c_V$, and the magnetization.
Let us first consider the former. We expand the free energy functional, Eq.\ 
(\ref{eq:7}), about the expectation value, $m$, of ${\bf M}$ to second
order, and then perform the Gaussian integral to obtain the partition
function. The free energy is obtained as the sum of
a mean-field contribution given by $\Phi[m]$, and a fluctuation contribution
given by the Gaussian integral. The latter yields the leading nonanalytic
term in the free energy.
We find \cite{ustbp} that effectively $H$ and $T$ have the 
same scale dimension, viz. $d (=z)$, and that at $t=0$ there is
a logarithmic $T$-dependence of $\gamma_V$ for {\it all} $2<d<4$.
These are the logarithmic corrections to 
scaling mentioned above. If we put the $t$-dependence back in, 
we obtain that the scale dependence of $\gamma_V$ is given by 
Eq.\ (\ref{eq:4}).

For the magnetization, the situation
is different compared to the specific heat, in that the leading $T$-dependence
is given by the mean-field or saddle-point contribution to the free energy.
We calculate the temperature corrections to the equation of state, Eq.\ 
(\ref{eq:2a}), and find that for $m>>T$ (in suitable units), $m^{d/2}$ in
Eq.\ (\ref{eq:2a}) will be replaced by 
$m^{d/2}[1+{\rm const}\times T/m + \ldots]$,
while for $m<<T$, $t$ is replaced by $t + T^{1/2\nu}$. The effective scale
dimension of $T$ in $m$ is therefore $2$ ({\it not} $z$), and we obtain
for $m$ the homogeneity law given by Eq.\ (\ref{eq:3}).
The appropriate interpretation of the relevant operator $T$
in these equations is that it reflects the crossover from the quantum FP
to the classical FP rather than dynamical scaling. 
Accordingly, we have written the $T$-dependence
in Eq.\ (\ref{eq:3}) in terms of the crossover exponent $\phi$.

We finally discuss some further aspects of our results, and their
relation to other work.
The phase transition described above
is very similar to the one discussed in Ref.\cite{IFS}. In particular,
the critical exponents on the disordered side and in the quantum
critical region are the same, and Ref.\ \cite{IFS} also found
logarithmic corrections to scaling.
We therefore believe that the quantum phase transition found in Ref.
\cite{IFS}, whose physical nature was unclear, was the ferromagnetic 
transition.
Also note that throughout we have studied what amounts to the average
effects of disorder; the coefficients in our LGW functional are disorder
averaged correlation functions. In addition, fluctuation effects appear,
e.g. in the form
of a random mass term at $O({\bf M}^4)$. The 
coefficient of
this term is the disorder average of the square of the unaveraged
correlation function $\chi^{(2)}$. This term is irrelevant by power
counting with respect to our FP for all $d$ except $d=4$, where it
is marginal, and all other fluctuation terms are also irrelevant\cite{ustbp}.

Sachdev \cite{Sachdev} has performed a very general
scaling analysis of quantum phase transitions with conserved order parameters.
Some of his results do not directly apply to our
problem because the coefficient $v$ in Eq.\ (\ref{eq:10}) acts as a 
dangerous irrelevant variable (DIV). For instance, it follows from
our Eq.\ (\ref{eq:3}) that the Wilson ratio,
$W = (m/H)/(c_V/T)$, diverges at criticality rather than being a universal
number. The reason for this breakdown of general scaling
is the flexibility displayed by DIV in modifying the scaling behavior of
various quantities: Not only can a particular irrelevant variable be
dangerous for some observables but not for others, it can also affect
different arguments of a given scaling function in different ways. Here,
the DIV $v$ affects the magnetization differently than the specific heat
coefficient, and in the latter it changes the effective scale dimension of $H$
(from $(3d-2)/2$ to $d$ in $2<d<4$), but not that of $T$. 

We gratefully acknowledge discussions with Subir Sachdev and W.W. Warren.
This work was supported by the NSF under grant numbers DMR-92-09879,
DMR-92-17496, and DMR-95-10185.

\end{document}